\documentclass[aps,prb,twocolumn,groupedaddress,showpacs,floatfix]{revtex4-1}

\usepackage{graphicx}
\usepackage{dcolumn}
\usepackage{bm}
\usepackage{amsmath}

\begin{document}

\title{Patterns of the Aharonov-Bohm oscillations in graphene nanorings}

\author{Igor Romanovsky}
\email{Igor.Romanovsky@physics.gatech.edu}
\author{Constantine Yannouleas}
\email{Constantine.Yannouleas@physics.gatech.edu}
\author{Uzi Landman}
\email{Uzi.Landman@physics.gatech.edu}

\affiliation{School of Physics, Georgia Institute of Technology,
             Atlanta, Georgia 30332-0430}

\date{2 December 2011; Physical Review B {\bf 85}, 165434 (2012)}

\begin{abstract}
Using extensive tight-binding calculations, we investigate (including the spin) the Aharonov-Bohm 
(AB) effect in monolayer and bilayer trigonal and hexagonal graphene rings with zigzag boundary 
conditions. Unlike the previous literature, we demonstrate the universality of integer ($hc/e$) 
and half-integer ($hc/2e$) values for the period of the AB oscillations as a function of the 
magnetic flux, in consonance with the case of mesoscopic metal rings. Odd-even (in the number of
Dirac electrons, $N$) sawtooth-type patterns relating to the halving of the period have also 
been found; they are more numerous for a monolayer hexagonal ring, compared to the cases of a 
trigonal and a bilayer hexagonal ring. Additional more complicated patterns are also present, 
depending on the shape of the graphene ring. Overall, the AB patterns repeat themselves as a 
function of $N$ with periods proportional to the number of the sides of the rings. 
\end{abstract}

\pacs{73.23.-b, 73.22.Pr, 73.23.Ra}

\maketitle

\section{Introduction}

Due to widespread interest in nanoscience and nanotechnology in the past fifteen years,
persistent currents (PCs) and the Aharonov-Bohm (AB) effect in ring-type nanosystems 
have attracted much attention. Originally, PCs and the AB effect were studied theoretically 
for spinless electrons in the ideal case of strictly one-dimensional (zero-width) metallic
nanorings threaded by a solenoidal magnetic flux. \cite{imry83,imry86,gefe88,glaz09} Subsequently,
consideration of spin in this ideal case was shown \cite{loss91} to lead to a nontrivial odd-even 
behavior, associated with halving ($\Phi_0/2$ versus $\Phi_0$) of the universal AB period and of 
the corresponding amplitude of the AB oscillations as a function of the applied magnetic
field $B$; $\Phi_0=hc/e$ is the unit of magnetic flux.

Recently fabricated carbon-based new materials, like carbon nanotubes \cite{roch01} and
two-dimensional graphene, provide additional opportunities for investigations of PCs and the AB 
effect, with potential future technological applications, in ring-type nanodevices. 
However, in spite of the recent extraordinary interest in graphene (starting with the isolation of
a single graphene sheet \cite{geim04}), only a few experimental \cite{russ08,ihn10} and 
theoretical studies (see, e.g., Refs.\ \onlinecite{rech07,wei10,wurm10,ma10}) 
of PCs and the AB effect in graphene nanorings have 
appeared in the last couple of years. Surprisingly these graphene-ring studies have been 
inconclusive regarding the aforementioned odd-even behavior associated with the electron spin; 
at the same time, no regular behavior or other pattern of the AB oscillations was reported.
Moreover, one \cite{wei10} of these publications has concluded that the odd-even behavior 
fails to manifest in graphene nanorings at all. 

In this letter, based on extensive tight-binding calculations, we investigate the AB 
oscillations for the case of trigonal and hexagonal narrow graphene rings terminating in zigzag 
edges; for experimental advances in the fabrication of graphene samples with well-defined 
high-purity edges, see Ref. \onlinecite{note2}. Our systematic studies (in the size range 
$1 \leq N \leq 100$ Dirac electrons) reveal clear signatures of several well defined patterns 
(including odd-even and halved-period behaviors) that can be traced to consideration of both the 
spin degree of freedom and of the zigzag boundary conditions obeyed by the graphene Dirac electrons. 
The different conclusion arrived in this Letter in comparison with previous publications 
\cite{rech07,wei10} appears to be due to the simplified \cite{fert10} condition (infinite-mass 
boundary condition, which, unlike the zigzag condition, cannot describe different crystallographic
terminations and corner geometries in graphene) used in the latter, in conjunction with the circular 
symmetry required for obtaining analytic solutions of the continuous Dirac-Weyl equation.

\begin{figure}[t]
\centering\includegraphics[width=6.8cm]{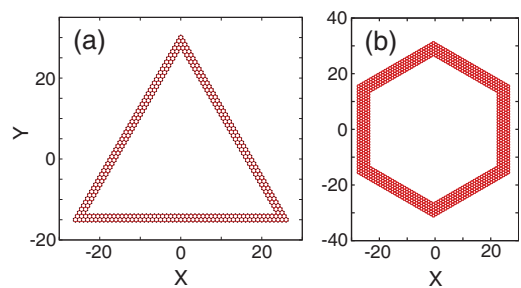}
\caption{
(Color online) Diagram of the narrow trigonal and hexagonal graphene rings with zigzag boundary 
conditions (for both the inner and outer edges) used in the TB calculations. (a) equilateral 
trigonal ring with a width of three rows of carbon atoms (b) hexagonal ring (with edges forming 
concentric regular polygons), with a width of five rows.
The length unit is the lattice constant $a=0.246$ nm.
}
\label{shape}
\end{figure}

\section{Preliminary theoretical background}

The spectra of an ideal metallic ring \cite{gefe88} (IMR) are very regular exhibiting a parabolic 
dependence on the magnetic flux $\Phi$, which is portrayed by the simple analytic expression
\begin{equation} 
\varepsilon^\text{IMR}_i(\Phi) \propto (l-\Phi/\Phi_0)^2,
\label{eidr}
\end{equation}
where the single-particle angular momentum $l$ takes the values $l=0, \pm 1, \pm 2,\dots$. 
This regularity is directly reflected in AB related 
quantities, such as the presistent current $I$ and the total magnetization $M$, which exhibit a 
periodic behavior as a function of $\Phi$ with period $\Phi_0$ (for spinless electrons 
\cite{gefe88}) or both $\Phi_0$ and $\Phi_0/2$ (when the electron spin is considered. 
\cite{loss91}) Indeed one has,

\begin{equation}
I = -c \frac{d E_{\text{tot}}}{d \Phi} \text{~~~~and~~~~}
M = - \frac{d {E_\text{tot}}}{d B},
\label{pc}
\end{equation}
where the total energy 
\begin{equation}
E_{\text{tot}}=\sum_{i,\sigma}^{\text{occ}} \varepsilon_i(B)
\label{etot}
\end{equation}
is given by the sum over all occupied single-particle (noninteracting 
electrons \cite{note3}) energies; the index $\sigma$ runs over spins. 
The magnetic flux in Eq.\ (\ref{pc}) is specified as $\Phi=BS$, 
where the area $S=\pi R^2$, with $R$ being the radius of the 1D ideal ring; for advances in the
measurement of small persistent currents and magnetic moments, see Ref.\ \onlinecite{note2}(b). 

To determine the single-particle spectrum (the energy levels $\varepsilon_i(B)$) in the 
tight-binding (TB) calculations for the graphene rings, we use the hamiltonian
\begin{equation}
H_{\text{TB}}= - \sum_{<i,j>} t_{ij} c^\dagger_i c_j + h.c.,
\label{htb}
\end{equation}
with $< >$ indicating summation over the nearest-neighbor sites $i,j$. The hopping
matrix element 
\begin{equation}
t_{ij}=t \exp \left( \frac{ie}{\hbar c}  \int_{{\bf r}_i}^{{\bf r}_j} 
d{\bf s} \cdot {\bf A} ({\bf r}) \right), 
\label{tpei}
\end{equation}
where $t=2.7$ eV, ${\bf r}_i$ and ${\bf r}_j$ are the positions of the carbon atoms
$i$ and $j$, respectively, and  ${\bf A}$ is the vector potential associated with the
applied perpendicular magnetic field $B$. The diagonalization of the TB hamiltonian [Eq.\ 
(\ref{htb})] is implemented with the use of the sparse-matrix solver ARPACK. \cite{arpack}  
In calculating $E_{\text{tot}}$ [see Eq.\ (\ref{pc})], only the single-particle TB energies with 
$\varepsilon_i(B) > 0$ are considered. \cite{rech07,wei10} We note here that, unlike the
continuous Dirac-Weyl equations, \cite{rech07,wei10} both the $K$ and $K^\prime$ valleys are 
automatically incorporated in the tight-binding treatment of graphene nanorings.

\begin{figure}[t]
\centering\includegraphics[width=6.2cm]{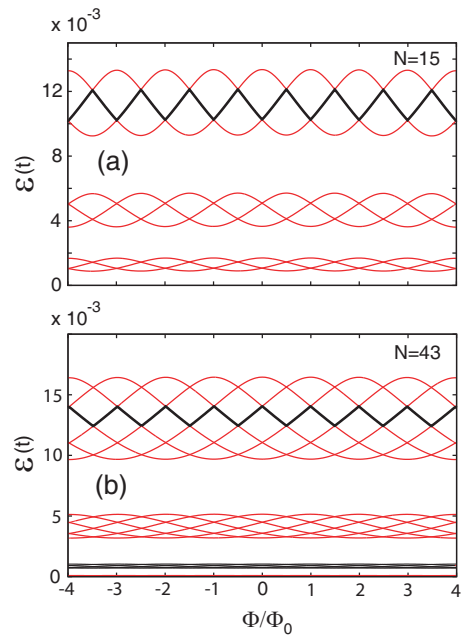}
\caption{
(Color online) Characteristic TB low-energy spectra of the narrow graphene rings with zigzag 
boundary conditions portayed in Fig.\ \ref{shape}. (a) trigonal graphene ring. (b) hexagonal
graphene ring. The thick black lines indicate the highest occupied state for $N=15$ [in (a)] and  
$N=43$ [in (b)] Dirac electrons (spin included).
Note the three-fold energy bands for the trigonal ring in (a) and the six-fold ones for the
hexagonal ring in (b). In the case of the trigonal ring (odd number of sides), the consecutive
three-fold bands are shifted by a phase $\Phi_0/2$ with respect of each other; this results to
a doubling of the period of the AB patterns as a function of $N$, i.e., a period of twelve 
instead of six (spin included). In the case of the hexagonal-ring spectrum (even number of
sides), no such shift is present, and the period as a function of $N$ remains twelve (spin
included).  
}
\label{spect}
\end{figure}

\section{Monolayer trigonal ring}
\label{secmt}

First we analyze TB results for a narrow trigonal graphene ring having pure zigzag terminations 
for both the inner and outer edges; see Fig.\ \ref{shape}(a). The corresponding TB spectra are 
displayed in Fig.\ \ref{spect}(a). Since the constant magnetic field $B$ is applied across 
the whole width of the ring, the magnetic flux is defined here in an average 
sense, i.e., through the use of an average area $S_{\text{av}}$ given by 
\begin{equation}
S_{\text{av}} \approx (S_{\text{inn}} + S_{\text{out}})/{2},
\label{avfl}
\end{equation}
where the indices ``inn'' and ``out'' indicate the areas enclosed by the 
inner and outer edges of the ring, respectively.  

\begin{figure}[t]
\centering\includegraphics[width=8.2cm]{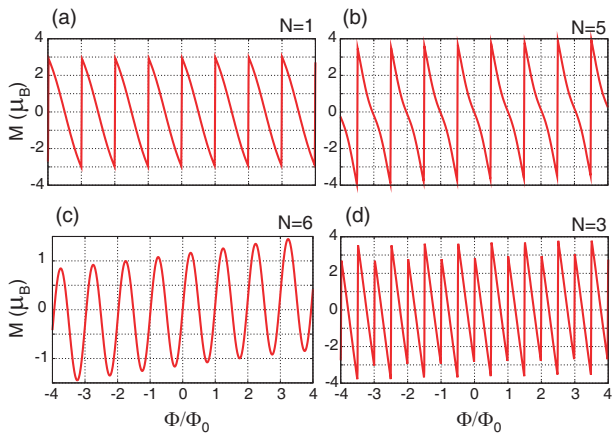}
\caption{
(Color online) Magnetization as a function of the the magnetic flux $\Phi$ (spin is included). The 
panels portray the four characteristic patterns of the Aharonov-Bohm oscillations associated with 
the trigonal graphene ring having zigzag boundary terminations; see Fig.\ \ref{shape}(a).
(a) Sawtooth. (b) Pinched sawtooth. (c) Asymmetric rounded sawtooth. (d) Halved-period sawtooth.
}
\label{mag1z}
\end{figure}
The graphene-ring spectra in Fig.\ \ref{spect}(a) are different from the simple 
spectra in Eq.\ (\ref{eidr}), familiar from the case of 1D metallic rings. \cite{gefe88}
Specifically, they are grouped in bunches of six levels (see also Ref.\ \onlinecite{baha09}), 
and each such bunch contains two three-level units. Naturally, this organization will be reflected 
in the behavior of the Aharonov-Bohm oscillations. Indeed, we found that the AB oscillations for the 
magnetization $M(\Phi)$ exhibits an overall period of $2\times6=12$ as a function of the electron 
number $N$ (the factor of 2 resulting from the spin degree of freedom). Within this period of 12
electrons, we find four distinct patterns as a function of $\Phi$ (see Fig.\ \ref{mag1z}), 
namely (a) sawtooth, (b) pinched sawtooth, (c) asymmetric rounded sawtooth, and (d) halved-period
sawtooth.

The first three patterns [Fig.\ \ref{mag1z}(a-c)] exhibit a period of $\Phi_0$ as a function of 
$\Phi$, while the fourth pattern [Fig.\ \ref{mag1z}(d)] has a halved period $\Phi_0/2$. 
As aforementioned, the halving of the fundamental period 
$\Phi_0$ was seen earlier in studies \cite{loss91} of the AB effect for spinfull electrons in 
ideal 1D metallic rings. In this case, it was described as an odd-even effect due to a 
two-electron alternation as a function of $N$. In contrast, the halving of the fundamental period
in the case of trigonal graphene nanorings exhibits a six-electron period as a function of $N$,
namely for $N=6i+N_0$, and only when $N_0=3$ ($i=1,2,\ldots$).

Another regular behavior in the AB patterns of trigonal graphene rings is a constant shift 
of the $\Phi$-dependence by $\pm \Phi_0/2$ for all electron sizes related by $N=6i+N_0$, with
$N_0=1,2, ..., 6$; $N_0$ is kept constant while $i$ runs over $i=1,2,3, \ldots$. For example,
the pattern of $N=8$ is the same as that of $N=2$, but shifted by $\Phi_0/2$, and the same holds
for the pattern of $N=10$ relative to that of $N=4$, etc.

Taking consideration of the above, and through inspection of magnetization curves in the range
$1 \leq N \leq 100$, the following summary of the AB patterns can be deduced ($i=1,2,\ldots$):\\
1. Sawtooth pattern (a) with zero shift: $N=12 i +1$, $N=12 i+2$, $N=12i +10$.\\
2. Sawtooth pattern (a) with a $\Phi_0/2$ shift: $N=12 i +4$, $N=12 i+7$, $N=12i +8$.\\
3. Pinched sawtooth pattern (b) with a $\Phi_0/2$ shift: $N=12 i +5$.\\
4. Pinched sawtooth pattern (b) with zero shift: $N=12 i +11$.\\
5. Asymmetric rounded sawtooth pattern (c) with zero shift: $N=12 i +6$.\\
6. Asymmetric rounded sawtooth pattern (c) with a $\Phi_0/2$ shift: $N=12 i +12$.\\
7. Halved-period sawtooth pattern (d) with zero shift: $N=12 i +3$.\\
8. Halved-period sawtooth pattern (d) with a $\Phi_0/2$ shift: $N=12 i +9$.

To summarize, $\Phi_0/2$-oscillations as a function of the magnetic flux occur only in cases
7 and 8 above, with the latter involving also an overall $\Phi_0/2$ shift.

\begin{figure}[t]
\centering\includegraphics[width=8.4cm]{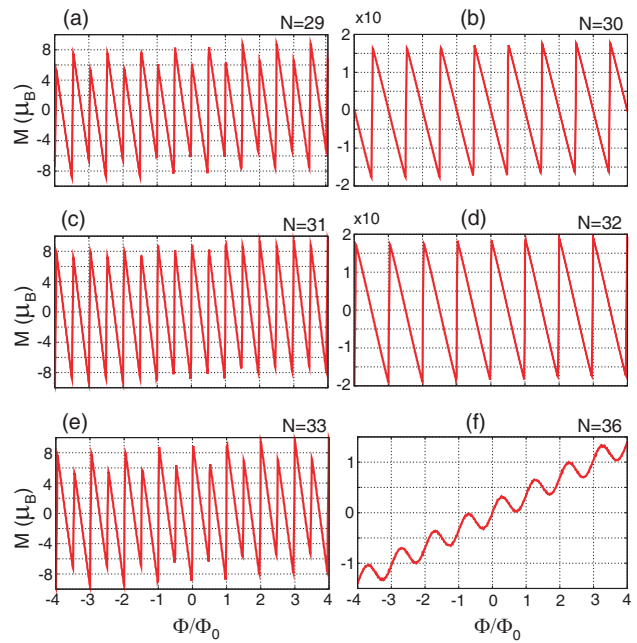}
\caption{
(Color online) Magnetization as a function of the the magnetic flux $\Phi$ (spin is included). The 
panels portray six patterns of the Aharonov-Bohm oscillations associated with the hexagonal 
graphene ring having zigzag boundary terminations [see Fig.\ \ref{shape}(b)]. The first five 
patterns [(a)-(e)] correspond to single-particle states near the middle of the 12-fold spectral 
band (spin included), while the sixth pattern corresponds to the top state [see Fig.\ 
\ref{spect}(b)]. (a) $N=29$; shifted halved-period sawtooth pattern. (b) $N=30$; shifted sawtooth.
(c) $N=31$; halved-period sawtooth pattern. (d) $N=32$; sawtooth pattern.
(e) $N=33$; halved-period sawtooth. (f) $N=36$; rounded sawtooth. 
The qualitative development in [(a)-(e)] of an odd-even alternation between one-period, $\Phi_0$, 
and halved-period, $\Phi_0/2$, sawtooth patterns is evident.
}
\label{hxmag0z}
\end{figure}

\section{Monolayer hexagonal ring}
\label{secmh}

Next we analyze the AB oscillations in the case of a narrow hexagonal graphene ring with zigzag
edges [see Fig.\ \ref{shape}(b)]. The corresponding energy spectrum [see Fig.\ \ref{spect}(b)] 
exhibits again an organization in bands, as was the case with the spectra of the trigonal ring.
However, each band now contains six, instead of three, single-particle levels, and this is clearly
connected to the sixfold point-group symmetry of the regular hexagon (the three-level bands arising
also from the threefold symmetry of the equilateral triangle).

Compared to the trigonal-ring spectra, the hexagonal-ring spectra are simpler in one way;
namely, there is no phase shift between two successive sixfold bands [see Fig.\ \ref{spect}(b)], 
in contrast to the $\Phi_0/2$ shift between successive threefold bands for the trigonal rings [see
Fig.\ \ref{spect}(a)]. The presence (absence) of a $\Phi_0/2$ shift between successive bands 
appears to be a general behavior of the spectra of regular-polygon-shaped graphene rings with 
odd (even) number of sides.

The absence of a shift between consecutive energy bands leads to a simplification of the
Aharonov-Bohm patterns, since it results in a period of $2\times 6=12$ (avoiding the doubling to 
24) electrons as a function of $N$. Of particulat interest is the fact that, disregarding a 
potential shift of $\pm \Phi_0/2$, the AB patterns exhibited by the magnetization curves (see 
Fig.\ \ref{hxmag0z}) display a well developed (although apparently not perfect) alternation 
pattern between integer periods $(\Phi_0)$ and halved periods $(\Phi_0/2)$, as long as the highest
occupied state lies in the interior of the sixfold energy band. The $\Phi_0/2$ period reflects
the zigzag nature of the interior states (which we term W-states to distinguish from the zigzag 
boundary condition); examples of W-states are given by the thick black lines in Fig.\ \ref{spect}.
When the Fermi level (highest occupied state) coincides with a W-state, $\Phi_0/2$-oscillations
occur. Note that there are four W-states for the hexagonal ring, but only one W-state for the 
trigonal ring. 

In Fig.\ \ref{hxmag0z}, we display the magnetization curves for several instances of electrons
occupying states in the 12-fold band with the number of electrons ranging from $N=25$ to $N=36$ 
(the doubling $2\times 6=12$ is due to consideration of the electron spin). In the range 
$27 \leq N \leq 34$, the magnetization curves exhibit an odd-even effect associated with the 
alternation between a whole-period ($\Phi_0$) sawtooth oscillation and a halved-period 
($\Phi_0/2$) sawtooth pattern {\it (exhibiting also a halved amplitude)\/}; examples of this 
behavior are portrayed in Fig.\ \ref{hxmag0z}[(a)-(e)]. The two cases for $N=25$ and $N=26$, with
the 25th and 26th electrons occupying the bottom level of the sixfold band, exhibit both a 
full-period ($\Phi_0$) sawtooth behavior. Finally, the two electrons occupying the top level 
of this energy band (corresponding to $N=35$ and $N=36$) exhibit a dissimilar behavior, with the 
penultimate one ($N=35$) having a full-period ($\Phi_0$) sawtooth behavior and the ultimate one 
($N=36$) showing a full-period ($\Phi_0$) rounded-sawtooth behavior [see Fig.\ \ref{hxmag0z}(f)]. 
Naturally, the aforementined AB patterns repeat themselves with a period of 12 electrons.

\begin{figure}[t]
\centering\includegraphics[width=8.0cm]{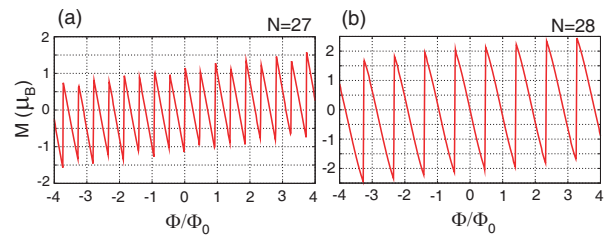}
\caption{
(Color online)  Patterns of the Aharonov-Bohm oscillations associated with a wider [in comparison
with Fig.\ \ref{shape}(b)] hexagonal monolayer graphene ring (with 12 rows of carbon atoms at each
side) having zigzag boundary terminations.
(a) $N=27$; halved-period sawtooth. (b) $N=28$; sawtooth.
}
\label{hxmag2z}
\end{figure}

In Fig.\ \ref{hxmag2z}, we display illustrative magnetization curves for the case of a wider 
hexagonal ring compared to the one in Fig.\ \ref{shape}(b) (by a factor of 2.4). From an
inspection of the patterns in Fig.\ \ref{hxmag2z}, as well as others not shown here, we found
that the behavior of the AB oscillations in this wider ring change only in minor ways. Much 
wider rings are needed to reach a substantial modification in the AB behavior. 

\begin{figure}[t]
\centering\includegraphics[width=6.5cm]{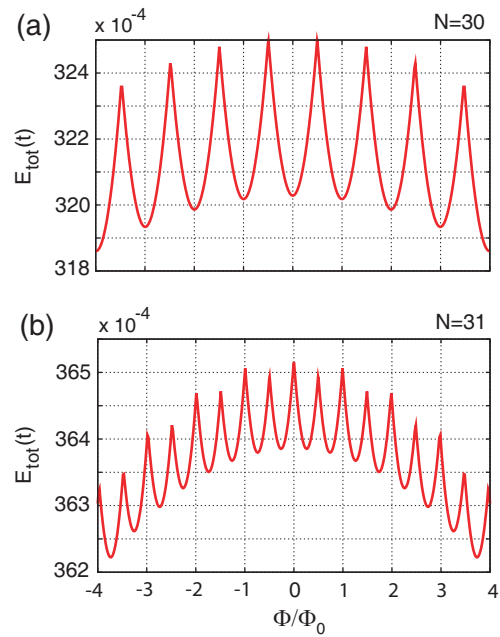}
\caption{
(Color online) Total energy curves (as a function of the magnetic flux $\Phi$) corresponding to 
the magnetizations in Figs.\ \ref{hxmag0z}(b) and \ref{hxmag0z}(c) [case of the thin monolayer 
hexagonal graphene ring with zigzag terminations portrayed in Fig.\ \ref{shape}(b)]. (a) $N=30$. 
(b) $N=31$. Observe the doubling of the frequency and the halving of the amplitude of the
oscillations as one goes from $N=30$ (even) to $N=31$ (odd).
}
\label{metot}
\end{figure}

\section{Similarities with the ideal metal ring}
\label{secsim}

To gain further insight into the appearance of the odd-even AB behavior in graphene nanorings with
zigzag terminations (described in Secs. \ref{secmt} and \ref{secmh}), we plot in Fig.\ \ref{metot}
the total energy curves, ${E_\text{tot}}(\Phi)$ [see Eq.\ (\ref{etot})], as a function of the 
average magnetic flux $\Phi$ [see Eq.\ (\ref{avfl})] for two characteristic cases; namely for $N=30$
and $N=31$ discussed earlier for an hexagonal graphene ring, see Fig.\ \ref{hxmag0z}(b) and 
\ref{hxmag0z}(c). 

A remarkable feature of these total energy curves is the almost parabolic ($\propto \Phi^2$) 
dependence on the magnetic flux (equivalently the applied magnetic field), which exhibit a period 
$\Phi_0$ for $N=30$ (even) and a half period $\Phi_0/2$ for $N=31$ (odd). The odd-even sawtooth 
oscillations of the magnetization portrayed in Fig.\ \ref{hxmag0z} are a direct consequence of this 
parabolic dependence given the definition of the magnetization as the derivative of the total energy
with respect to the magnetic flux [see Eq.\ (\ref{pc})]. 

We have further examined the total energy curves, ${E^\text{IMR}_\text{tot}}(\Phi)$ (not shown 
here), for the case of an ideal metallic ring, i.e., using the well known analytic energies of Eq.\
(\ref{eidr}), and have confirmed that their shape consists of similar parabolic segments 
exhibiting a $\Phi_0$ or a $\Phi_0/2$ period for even or odd $N$, respectively. 

Naturally, this overall parabolic ($\propto \Phi^2$) dependence of ${E^\text{IMR}_\text{tot}}(\Phi)$
could have been anticipated due to the original parabolic dependence on $\Phi$ of the single-particle
levels $\varepsilon^\text{IMR}_i(\Phi)$ [see Eq.\ (\ref{eidr})]. However, for graphene rings
with zigzag terminations, this result is a surprising one, given that the
associated single-particle spectrum is much more complicated; it further indicates that
the corresponding graphene single-particle energies [associated with the W-states, see 
Secs. \ref{secmt} and \ref{secmh}] are parabolic on $\Phi$ to a rather large degree.
  
We further briefly mention here that in preliminary calculations we found that graphene
rings with armchair edge terminations have, in contrast to those with zigzag terminations, 
single-particle spectra with an almost linear dependence on $\Phi$, and thus their AB patterns are
different (as we will describe in detail elsewhere \cite{igor12}).

\begin{figure}[t]
\centering\includegraphics[width=6.5cm]{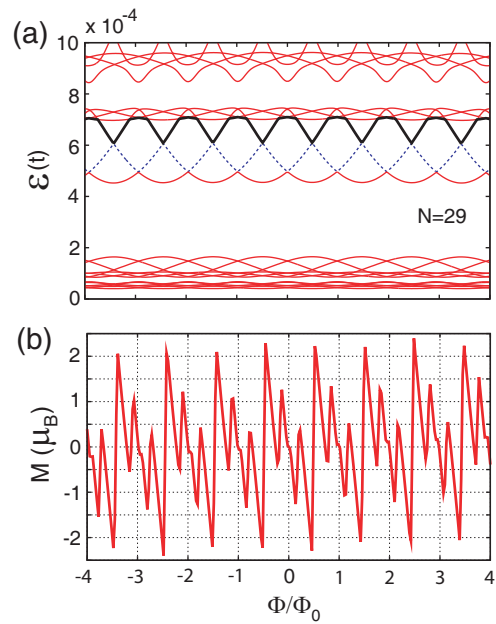}
\caption{
(Color online) An example for the case of a narrow bilayer hexagonal graphene ring with zigzag 
terminations in all twelve sides. The bilayer ring is built by stacking (Bernal stacking) two
monolayer hexagonal rings resembling the shape in Fig.\ \ref{shape}(b). (a) Characteristic
part of the spectrum. The thick black line denotes the level occupied by the 29th electron
(spin included). The dashed line (online blue) denotes the single W-state here.
(b) Corresponding magnetization curve (for $N=29$ electrons) as a function of 
the the magnetic flux $\Phi$ (spin is included). 
}
\label{bi-hxmag0z}
\end{figure}

\section{Bilayer hexagonal ring}

Having addressed the appearance of regular trends in the AB oscillations of monolayer graphene 
nanorings, we comment next on possible modifications that arise in associated bilayer graphene-ring
structures. To this end, we consider an hexagonal bilayer ring formed by stacking two monolayer
rings [resembling the arrangement portrayed in Fig.\ \ref{shape}(b)] one on top of the other 
following the Bernal prescription. Due to the Bernal-type coupling between the two rings, such 
narrow bilayer graphene rings are analogs of the double-ring configurations considered recently
in the framework of the Aharonov-Bohm effect in mesoscopic metallic devices. \cite{avis09}

A charecteristic part of the low-energy TB spectra for the bilayer ring is displayed in Fig.\
\ref{bi-hxmag0z}(a). As was the case with the monolayer hexagonal rings, the emergence of sixfold 
energy bands persists also for the case of a narrow bilayer hexagonal ring. However, the couplings
between the layers leads to strong modifications within each energy band; namely, the three top 
energy levels are strongly compressed compared to the three bottom ones. This results in turn in 
several more complicated profiles for the AB oscillations, an example of which is displayed in 
Fig.\ \ref{bi-hxmag0z}(b). From an inspection of Fig.\ \ref{bi-hxmag0z}(a), it is also clear that 
there is only a single well-formed W-state that may serve as a Fermi level [see second level from
the bottom denoted by a dashed line (online blue)], and thus a halved-period sawtooth pattern 
occurs only once within the period of twelve electrons (with the spin degeneracy being accounted 
for).

\section{Conclusions} 

Using TB calculations and taking into account the spin, we have demonstrated 
the universality of the integer $(\Phi_0)$ and half-integer $(\Phi_0/2)$ magnetic-flux periods in 
the Aharonov-Bohm effect in narrow graphene rings with zigzag boundary conditions (trigonal and
hexagonal shapes were considered in both monolayer and bilayer structures). The AB patterns for
the monolayer hexagonal rings are dominated by an odd-even (in the electron number) alternation
of sawtooth-type oscillations with $\Phi_0$ and $\Phi_0/2$ periods. Such an odd-even alternation 
persists also for trigonal monolayer and hexagonal bilayer rings, with a reduced occurrence
frequency (related to the number of W-states in each energy band). Additional patterns of
higher complexity are also prominent, depending on the structure of the graphene ring. All AB 
patterns repeat themselves as a function of $N$ with periods relating to the point-group symmetry 
of the geometrical shape of the rings. \cite{note} Our findings, which contrast with the results 
of recent literature on the subject (see, e.g., Refs.\ \onlinecite{rech07,wei10}), provide the 
impetus for experimental probing of the AB effects in the graphene systems explored in this paper.

\acknowledgments

This work was supported by the Office of Basic Energy Sciences of the US D.O.E.
under contract FG05-86ER45234.


\begin{thebibliography}{999}
\bibitem{imry83}
M. B\"{u}ttiker, Y. Imry, and R. Landauer,
Phys. Lett. {\bf 96A}, 365 (1983).
\bibitem{imry86}
Y. Imry, in {\it Directions in Condensed Matter Physics\/},
edited by G. Grinstein and G. Mazenko 
(World Scientific, Singapore, 1986), p. 101.
\bibitem{gefe88}
Ho-Fai Cheung, Y. Gefen, E. K. Riedel, and Wei-Heng Shih,
Phys. Rev. B {\bf 37}, 6050 (1988).
\bibitem{glaz09}
A. C. Bleszynski-Jayich, W. E. Shanks, B. Peaudecerf, E. Ginossar, F. von Oppen,
L. Glazman, and J. G. E. Harris,
Science 326, 272 (2009).
\bibitem{loss91}
D. Loss and P. Goldbart,
Phys. Rev. B {\bf 43}, 13762 (1991).
\bibitem{roch01}
S. Roche {\it et al.\/},
Phys. Rev. B {\bf 64}, 121401(R) (2001); 
A. Bachtold {\it et al.\/}, 
Nature (London) {\bf 397}, 673 (1999).
\bibitem{geim04}
K. S. Novoselov {\it et al.\/},
Science {\bf 306}, 666 (2004).
\bibitem{russ08}
S. Russo, J. B. Oostinga, D. Wehenkel, H. B. Heersche, S. S. Sobhani, 
L. M. K. Vandersypen, and A. F. Morpurgo,
Phys. Rev. B {\bf 77}, 085413 (2008).   
\bibitem{ihn10}
M. Huefner, F. Molitor, A. Jacobsen, A. Pioda, Ch. Stampfer, K. Ensslin, and Th. Ihn,
New J. Phys. {\bf 12}, 043054 (2010).
\bibitem{rech07}
P. Recher, B. Trauzettel, A. Rycerz, Y. M. Blanter, C. W. J. Beenakker, and A. F. Morpurgo,
Phys. Rev. B {\bf 76}, 235404 (2007).
\bibitem{wei10}
C-H Yan and L-F Wei,
J. Phys.: Condens. Matter {\bf 22}, 295503 (2010).
\bibitem{wurm10}
J. Wurm, M. Wimmer, H. U. Baranger, and K. Richter,
Semicond. Sci. Technol. {\bf 25}, 034003 (2010). 
\bibitem{ma10}
M. M. Ma and J. W. Ding,
Solid State Commun. {\bf 150}, 1196 (2010).
\bibitem{note2}
Our studies have been motivated by the recent experimental advances concerning: (a) the 
fabrication and engineering of graphene edges with high-purity zigzag terminations; see, e.g.,
X. Jia {\it et al.\/}, Science {\bf 323}, 1701 (2009); B. Krauss {\it et al.\/}, Nano Lett. 
{\bf 10}, 4544 (2010); P. Nemes-Incze {\it et al.\/}, Nano Res. {\bf 3}, 110 (2010); R. Yang 
{\it et al.\/}, Adv. Mater. {\bf 22}, 4014 (2010); Zh. Shi {\it et al.\/}, Adv. Mater. {\bf 23}, 
3061 (2011); J. Lu {\it et al.\/}, Nature Nanotechnology {\bf 6}, 247 (2011). For the
engineering of high-purity armchair edges, see M. Begliarbekov {\it et al.\/}, Nano Lett. 
{\bf 11}, 4874 (2011).
(b) the ability to measure very small magnetic moments and currents; see Ref.\ \onlinecite{glaz09} 
and H. Bluhm {\it et al.\/}, Phys. Rev. Lett. {\bf 102}, 136802 (2009). For a perspective, see
Y. Imry, Physics {\bf 2}, 24 (2009).
\bibitem{fert10}
T. Luo, A. P. Iyengar, H. A. Fertig, and L. Brey,
Phys. Rev. B {\bf 80}, 165310 (2009);
H. A. Fertig and L. Brey,
Phil. Trans. R. Soc. A {\bf 368}, 5483 (2010).
\bibitem{note3}
While many-body effects may be of interest in the context of the AB/PC effect for a certain choice 
of parameters, e.g., when a Wigner molecule is formed [see, e.g., R. Okuyama, M. Eto, and H. Hyuga,
Phys. Rev. B {\bf 83}, 195311 (2011)], in the current  paper we focus on the broad range of 
instances where the noninteracting electron model provides an appropriate description. In this 
context, see the experimental study in Ref.\ \onlinecite{glaz09}, where for metal nanorings it was 
found that ``Measurements of both a single ring and arrays of rings agree well with calculations 
based on a model of non-interacting electrons.'' The results of a recent sole study of many-body 
effects in graphene rings [D. S. L. Abergel {\it et al.\/}, Phys. Rev. B {\bf  78}, 193405 (2008)] 
were obtained for convenience with the use of the simplified \cite{fert10} infinite-mass boundary 
condition, and consequently are not considered by us here. 
\bibitem{arpack}
R. B. Lehoucq, D. C. Sorensen, and C. Yang,
{\it ARPACK Users' Guide: Solution of Large-Scale Eigenvalue
Problems with Implicitly Restarted Arnoldi Methods\/} (SIAM,
Philadelphia, 1998).
\bibitem{baha09}
D. A. Bahamon, A. L. C. Pereira, and P. A. Schulz,
Phys. Rev. B {\bf 79}, 125414 (2009); see also Ref.\ \onlinecite{rech07}.
\bibitem{igor12}
I. Romanovsky, C. Yannouleas, and U. Landman,
to be published,
\bibitem{avis09}
Y. Avishai and J. M. Luck,
J. Phys. A: Math. Theor. {\bf 42}, 175301 (2009).
\bibitem{note}
These patterns are robust with respect to variations in the width of the rings (see, e.g.,
Fig.\ \ref{hxmag2z}), as well as to variations in their shape away from a regular polygon 
(see Ref.\ \onlinecite{igor12}).
\end{thebibliography}
\end{document}